# Renormalization of the Lattice Heavy Quark Classical Velocity

Jeffrey E. Mandula[a] and Michael C. Ogilvie[b]

[a]Department of Energy, Division of High Energy Physics
Washington, DC 20585

[b]Department of Physics, Washington University
St. Louis, MO 63130

In the lattice formulation of the Heavy Quark Effective Theory (LHQET), the "classical velocity" $v$ becomes renormalized. The origin of this renormalization is the reduction of Lorentz (or O(4)) invariance to (hyper)cubic invariance. The renormalization is finite and depends on the form of the decretization of the reduced heavy quark Dirac equation. For the Forward Time - Centered Space discretization, the renormalization is computed both perturbatively, to one loop, and non-perturbatively using two ensembles of lattices, one at $\beta = 5.7$ and the other at $\beta = 6.1$ The estimates agree, and indicate that for small classical velocities, $\vec{v}$ is reduced by about 25-30%.

## 1. INTRODUCTION

In the heavy quark limit[1,2] a single form factor, the Isgur-Wise universal function $\xi$, describes all semileptonic decays of one meson containing a heavy quark into another, such as the process $B \to D^* l \nu$. Its calculation is well suited to lattice gauge theory[3], and such calculations have been carried out using a lattice implementation of the heavy quark effective theory[4], also by treating the heavy quarks as Wilson fermions with a small hopping constant, without implementing the heavy quark limit[5,6].

On the lattice[4], as in the continuum[1,2,7], the Isgur-Wise limit entails the introduction of a "classical velocity", $v$, normalized to 1, which appears both in the decomposition of the momentum of a heavy particle and the reduced Dirac equation of the heavy quark field.

$$P = Mv + p$$

$$-iv \cdot D\, h^{(v)}(x) = 0 \quad (1)$$

In the continuum, the second relation is derived from the first, and the velocity that appears in these two contexts is the same. However, on the lattice this is not the case.

In this talk, we explain that the origin of this difference is the reduction of the space-time symmetry group from Lorentz (or Euclidean O(4)) invariance to hypercubic lattice symmetry. We then describe two calculations of its magnitude. The first is a one-loop perturbative calculation, which follows the analysis of Aglietti[8] (who however used a different discretization of the lattice Dirac operator than that adopted here). The second calculation is non-perturbative. It is based on a computation of the shift in the energy of a meson containing a heavy quark, as measured by the fall-off of its Euclidean space propagator, for a given shift in its residual momentum. In both calculations, it is useful to expand the shift in $v$ in a power series in the "bare" classical velocity, and in the conclusions we compare the first three terms calculated using each procedure.

## 2. ORIGIN OF THE RENORMALIZATION

The heavy quark effective theory is formulated[5] by factoring the $M \to \infty$ singular behavior from the field of a heavy quark, leaving a reduced operator $h^{(v)}$.

$$h^{(v)}(x) = e^{-iMv \cdot x} \frac{1 + \gamma \cdot v}{2} \psi(x) \quad (2)$$

The Lagrangian for $h^{(v)}$ is

$$\mathcal{L}^{(v)} = \bar{h}^{(v)}(x)\, iv \cdot D\, h^{(v)}(x) \quad (3)$$

and its propagator satisfies the reduced Dirac equation

$$-iv \cdot D\, S^{(v)} = \delta \quad (4)$$

Here $D$ is the covariant derivative. This structure is the same both on the lattice and in the continuum.

In the continuum, the propagator of the reduced field $h^{(v)}(x)$ is

$$S^{(v)}(p) = \frac{1}{v \cdot p - \Sigma^{(v)}(p)} \quad (5)$$

where $\Sigma^{(v)}(p)$ is the proper self mass. In perturbation theory, it is the sum of all the 1-particle irreducible self-mass diagrams.

If $\Sigma^{(v)}(p)$ is Taylor expanded in the residual momentum,

$$\Sigma^{(v)}(p) = m + \left.\frac{\partial \Sigma^{(v)}}{\partial p_\mu}\right|_{p=0} p_\mu + \cdots \quad (6)$$

the first term gives a mass shift. This is without physical significance and can be eliminated by a redefinition of the reduced heavy quark field by a phase which is independent of the heavy mass[9]. The renormalized classical velocity is inferred from the second term in the expansion. For small residual momentum, the full quark propagator has the behavior

$$S^{(v)}(p) = \frac{Z^{(v)}}{-m + v^{(ren)} \cdot p + \cdots} \quad (7)$$

This identifies the physical classical velocity as

$$v_\mu^{(ren)} = Z^{(v)}(v_\mu - X_\mu) \quad (8)$$

where

$$X_\mu \equiv \left.\frac{\partial \Sigma^{(v)}}{\partial p_\mu}\right|_{p=0} \quad (9)$$

The wave function renormalization constant $Z^{(v)}$ is determined by the requirement that both the bare and the renormalized classical velocities have unit normalization.

$$v_\mu^{(ren)2} = v_\mu^2 = Z^{(v)2}(v_\mu - X_\mu)^2 = -1 \quad (10)$$

On the lattice, $p$ is replaced by an appropriate discretization. This is the starting point of the calculation in perturbation theory.

The only 4-vector on which $X_\mu$ can depend is the bare classical velocity. In the continuum, the only possible dependence is for is to be linearly proportional to $v_\mu$. Therefore, it modifies the inverse of the heavy quark propagator to first order in $p$ only by an overall multiplicative factor, which is a wave function renormalization of $h^{(v)}(x)$, but not a change in the classical velocity.

On the lattice, because of the reduced Lorentz or rotational symmetry, $X_\mu$ need not be linearly proportional to $v_\mu$. For example, it can have a term proportional to $v_\mu^3$, which has the same hypercubic group transformation properties as $v_\mu$, but is linearly independent of it. When the coefficient $X_\mu$ is not simply proportional to $v_\mu$, the classical velocity is shifted.

## 3. NON-PERTURBATIVE SHIFT

A procedure for calculating the renormalized classical velocity which is suited to non-perturbative simulation is based on the observation that the rate of fall-off of the reduced propagator in Euclidean time for fixed residual spacial momentum is given by the residual energy at that momentum.

$$S(t,\vec{p}) \sim Z^{(v)} e^{-E^{(v)}(\vec{p})\, t} \quad (11)$$

The residual energy, which is directly amenable to simulation, has a simple dependence on the physical classical velocity and the residual momentum. In the $M \to \infty$ limit, the shift in the energy is

$$E^{(v)}(\vec{p}) - E^{(v)}(0) = \tilde{v}^{(phys)} \cdot \vec{p} \quad (12)$$

where the conventional classical energy (which is bounded by 1) is

$$\tilde{v}^{(phys)} = \frac{\vec{v}^{(phys)}}{v_0^{(phys)}} = \frac{\vec{v}^{(phys)}}{\sqrt{1 + \vec{v}^{(phys)2}}} \quad (13)$$

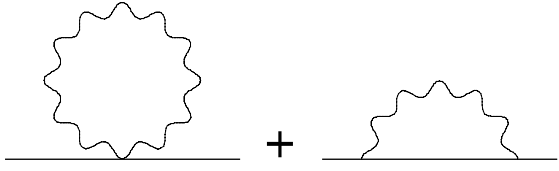

Figure 1 — The 1-loop contribution to the heavy quark proper self mass

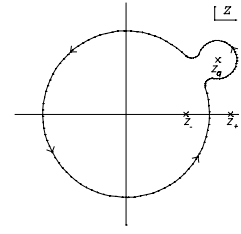

Figure 2 — Contour in the $z = e^{iEa}$ plane for the one loop proper self mass

For both the calculations, we employ the forward-time symmetrical-space discretization of the reduced Dirac equation

$$[U(x,x+\hat{t})\,S(x+\hat{t},y) - S(x,y)]$$
$$+ \sum_{\mu=1}^{3} \frac{-i\tilde{v}_\mu}{2} [U(x,x+\hat{\mu})\,S(x+\hat{\mu},y) \quad (14)$$
$$- U(x,x-\hat{\mu})\,S(x-\hat{\mu},y)]$$
$$= \delta(x,y)$$

The use of an asymmetric time difference is required in order to implement the requirement that heavy quarks propagate only forward in time.

The free inverse propagator corresponding to this discretization is

$$S^{(0)-1}(p) = v_0(e^{ip_4} - 1) + \sum_{\mu=1}^{3} v_\mu \sin p_\mu \quad (15)$$

As is evident from Eq. (14), the simulated propagator is naturally expanded in a power series in the (bare) classical velocity, which plays the role of a transverse hopping constant. To facilitate comparison of the perturbative and lattice calculations, we will also express the results of the perturbative calculation in powers of the bare classical velocity.

## 3. PERTURBATIVE EVALUATION

To one loop, the proper self mass is given by the two diagrams shown in Figure 1. The point interaction only gives rise to a residual mass, and so can be ignored. The second diagram does change the classical velocity, however. Because of the finite lattice spacing, the integration domain is periodic in the Euclidean 4-momentum. The heavy quark propagator is as in Eq. (15).

The evaluation of the loop integral over the gluon 4-momentum requires that the contour in the complex Euclidean energy plane be chosen so that the heavy quark propagator always vanishes for negative Euclidean time. This requires that in the $z$ plane it encloses the pole coming from the heavy quark propagator, as is indicated in Figure 2.

The lowest order perturbative renormalization

$$\delta\tilde{v}_i \equiv \tilde{v}_i^{(ren)} - \tilde{v}_i = -\frac{1}{v_0}(X_i - \tilde{v}_i X_0) \quad (16)$$

has the structure ($C_2$ is the quadratic Casimir)

$$\delta\tilde{v}_i = -g^2 C_2 [c_1 \tilde{v}_i + c_3 \tilde{v}_i^3 + c_{12} \tilde{v}_i \sum_{j \neq i} \tilde{v}_j^2 + \cdots] \quad (17)$$

The expansion coefficients are evaluated numerically:

$$\begin{aligned} c_1 &= .17800996 \\ c_3 &= .03159228 \\ c_{12} &= .02959309 \end{aligned} \quad (18)$$

## 4. SIMULATION

To avoid the ambiguities associated with global gauge fixing, we simulate the propagator of a heavy-light meson. Its classical velocity is the same as its heavy quark component. We have simulated the propagator expansion coefficients on an ensemble of lattices and Wilson light quark propagators made available to us by the Fermilab ACP-MAPS Collaboration[10]. These consisted of 48 lattices of size $24^3 \times 48$ with lattice coupling $\beta = 6.1$ along with Wilson quark propagators with hopping constant $\kappa = .154$. We computed the coefficients of the leading terms in the $\tilde{v}_i$ expansion of the heavy quark propagator, and formed heavy-light meson propagators expansion coefficients.

In this preliminary analysis, we did not optimize the wave function of the heavy-light meson, and in fact there was no extended region in Euclidean time over which the ground state was cleanly isolated. With that caveat, the propagator terms from which the physical classical velocity was extracted is shown in Figure 3.

## 5. COMPARISON OF RESULTS

The perturbative and simulation calculations of the coefficients of the first three terms in the bare classical velocity expansion of the finite renormalization of classical velocity is shown in the following Table:

Table I — Comparison of Simulated and Perturbative Expansion Coefficients

| Term | Simulation | Naive Perturb | Tadpole Improv |
|---|---|---|---|
| $v_i$ | $-.287 \pm .092$ | $-.2334$ | $-.3969$ |
| $v_i^3$ | $-.982 \pm .087$ | $-.0414$ | $-.0704$ |
| $v_i v_j^2$ ($i \neq j$) | $-.764 \pm .078$ | $-.0461$ | $-.0783$ |

The coefficient of the linear term agrees rather well between the simulated and perturbative evaluations. There is a statistically insignificant preference for the use of the naive rather than the tadpole improved coupling in this context. However, the perturbative and simulated cubic coefficients are in complete disagreement While it is possible that the lattice coupling is simply too large to neglect higher order terms, is should be pointed out that each of the cubic coefficients is the difference of two terms, each of which is substantially larger than their difference. Furthermore, each of the two terms has a $t^3$ asymptotic behavior, while their difference should only grow linearly. The large cancellations are roughly of the same magnitude in both cubic coefficients. Fortunately, the dominant, linear coefficient is free of this difficulty.

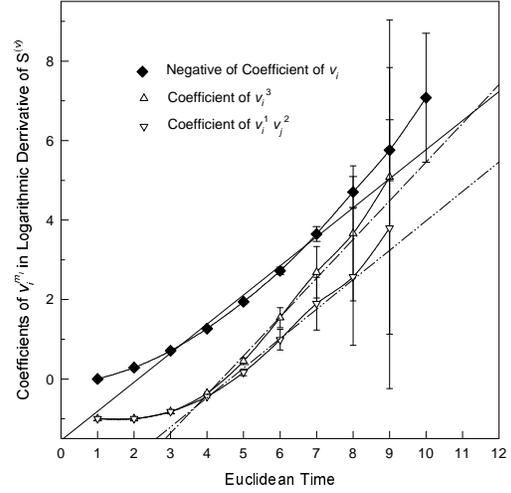

Figure 3 — Coefficients of the leading terms in the logarithmic derivative of $S^{(v)}$

Both the simulation and perturbation theory indicate that the physical classical velocity is smaller than its bare value.